\documentclass[twocolumn,showpacs,preprintnumbers,amsmath,amssymb]{revtex4}
\usepackage{dcolumn}
\usepackage{bm}
\usepackage{latexsym}
\usepackage{amsfonts}
\usepackage[dvips]{graphicx}
\usepackage[usenames]{color}
\usepackage{makeidx}

\newcommand{\pa}{\partial}

\newcommand{\ket}{\rangle }
\newcommand{\bra}{\langle }

\newcommand{\ve}{\varepsilon}

\newcommand{\up}{\uparrow}
\newcommand{\dw}{\downarrow}

\begin{document}

\title{
Photo-induced Tomonaga-Luttinger-like liquid in a Mott insulator}
\author{Takashi Oka and Hideo Aoki}
\address{Department of Physics, University of Tokyo, Hongo, Tokyo 113-0033, 
Japan}
\date{\today}
\begin{abstract}
\noindent 
Photo-induced metallic states in a Mott insulator 
are studied for the half-filled, one-dimensional Hubbard model 
with the time-dependent density matrix renormalization group.
An irradiation of strong AC fields is found to create, 
in the nonequilibrium steady state, a linear dispersion in the optical spectrum (current-current correlation) reminiscent of the 
Tomonaga-Luttinger liquid for the doped Mott insulator in equilibrium.  
The spin spectrum in nonequilibrium retains 
the des Cloizeaux-Pearson mode with the spin velocity 
differing from the charge velocity. 
The mechanism of the photocarrier-doping, along with 
the renormalization in the charge velocity, is analyzed 
in terms of an effective Dirac model. 
\end{abstract}

\maketitle

\maketitle
{\it Introduction ---} 
Doped one-dimensional(1d) Mott insulators 
are fascinating due to a special metallic state known as the 
Tomonaga-Luttinger (TL) liquid, where excitations have 
collective nature as distinct from the conventional 
Fermi liquid\cite{Schulz1992}. 
Specifically, the charge velocity becomes 
renormalized due to the electron-electron interaction, so that 
the charge excitation 
propagates with a velocity different from that of spin 
excitations --- a hallmark of the ``spin-charge separation". 
Experimentally, TL liquids have been observed in quantum wires 
and carbon nanotubes\cite{nanotubeLuttinger03}. 

Now, there is another way of making a Mott insulator metallic, 
which is entirely different from the chemical doping.  
Namely, photo-doping is now being highlighted as a way to 
control the carrier density in 1d correlated systems
\cite{Yu1991,Iwai2003,Okamoto2007,TakahashiGomiAihara02PRL,Matsueda2004,Maeshima05,Takahashi08}.  
Pump-probe experiments have shown that strong electric 
fields can turn Mott insulating crystals into 
metals\cite{Yu1991,Iwai2003,Okamoto2007}. 
Theoretically, the change of the conductivity by irradiation 
was studied in \cite{Maeshima05,Takahashi08}
with exact diagonalization.
Features in the photo-doping conceptually 
distinct from the chemical doping are: 
(a) the system is out of equilibrium, and 
(b) two types of carriers, i.e., electrons ($=$doubly occupied sites) 
and holes, coexist because they are pair-produced. 
Although electron-hole systems in equilibrium have been 
studied in the past in terms of the TL theory\cite{NagaosaOgawa}, 
properties, especially the spectral properties, of the {\it nonequilibrium} metallic states 
induced by irradiation are yet to be understood.

This has motivated the present work, where 
we shed light to this problem by studying the optical (current-current) 
and spin correlation functions for 
a nonequilibrium steady state in 
1d Mott insulators in strong AC electric fields. This is done in 
two steps: we first employ 
numerical simulations with the 
time-dependent density matrix renormalization group (td-DMRG)
\cite{tddmrg} 
to show that we have a TL-like linear dispersion 
in the photo-doped system, where 
the charge and spin excitations have 
different velocities. 
We then confirm the result from a field theoretic result.  
A starting point is ref.\cite{Oka2005a,Oka_LZreview}, 
where the present authors pointed out that there is an 
interesting analogy between 
the Schwinger mechanism in the decay of the 
quantum electrodynamics (QED) vacuum governed by the 
electron-positron creation rate\cite{Schwinger1951}, 
and the dielectric breakdown of Mott insulators 
in DC electric fields governed by the 
electron-hole creation rate. 
In 1d, correlated electron systems can be mapped 
to a simple Dirac model via the massive Thirring model, which 
is dual to the sine-Gordon model for charge degrees of freedom of the 
1d Hubbard model.  We then analyze the photo-doping 
with the effective Dirac model. 
There, the optical spectrum is formulated with the 
Floquet analysis for strong AC external fields.  
The result reproduces the main features in the td-DMRG result, providing a 
physical picture for the dynamics of the nonequilibrium collective 
modes in the photo-induced state.



{\it Nonequilibrium steady state ---}
We consider a Mott insulator in strong AC electric fields 
in the half-filled, 1d Hubbard model with the Hamiltonian 
given, in standard notation, by
$H(t)=H_0+H_{\rm F}(t)$, where 
$H_0=-t_{\rm hop}\sum_{i\sigma}(c_{i+1\sigma}^\dagger
c_{i\sigma}+\mbox{h.c.})+U\sum_i
n_{i\up}n_{i\dw}
$, 
and 
$
H_{\rm F}(t)=F\theta(t)\sin(\Omega t)\sum_iin_i$
($n_{i\sigma}=c_{i\sigma}^\dagger c_{i\sigma},\;
n_i=n_{i\up}+n_{\i\dw}$). 
Here $F$ and $\Omega$ are, respectively, the strength and 
frequency of the external electric field, 
which is switched on at $t=0$ (hence the insertion of a step function).
In the calculation we take the length of the system $L=80$, 
a time step $\Delta t=0.04$, 
and the DMRG Hilbert space size of $m=140$, in 
natural units. 
After obtaining the groundstate $|\Psi_0\ket$ of $H(t<0)$ 
with the finite-system DMRG algorithm \cite{dmrg}, 
we let the system evolve according to the time-dependent 
Hamiltonian $H(t)$ 
with the td-DMRG\cite{tddmrg} 
to obtain the wave function 
$|\Psi(t)\ket=U(t;0)|\Psi_0\ket$, 
with the time-evolution operator
$U(t;t')=\hat{T}e^{-i\int^t_{t'}H(s)ds}$
($\hat{T}$: the time-ordering). 

\begin{figure}[t]
\centering 
\includegraphics[width=7.5cm]{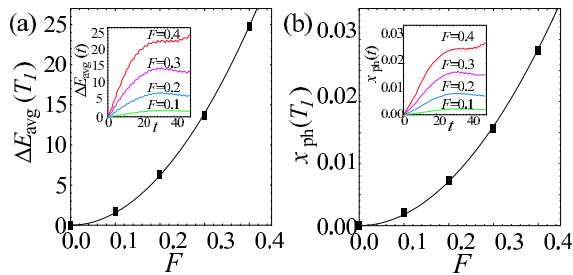}
\caption{
(Color online) Inset: 
Time-dependent DMRG result for 
the total energy(a), photo-doping rate(b) 
for the half-filled 1d Hubbard model with $U/t_{\rm hop}=8.0$, $\Omega=8.0$. 
The main panels depict their final values at $t=T_1$ vs 
the field strength $F$, where the solid lines are parabola fitted to the data.
}
\label{fig:energy}
\end{figure}

For $t>0$, the system relaxes into a nonequilibrium
steady state, where we can define the photo-doping rate as
$
x_{\rm ph}(t)=\frac{1}{L}\sum_i\left({\rm avg}\bra \Psi(t)|n_{i\up}n_{i\dw}|\Psi(t)\ket
-\bra \Psi_0|n_{i\up}n_{i\dw}|\Psi_0\ket\right)
$
which is the increment in the double occupancy.  
We also monitor the total energy, 
$E_{\rm avg}(t)= {\rm avg}\bra \Psi(t)|H(t)|\Psi(t)\ket$.  
In these expressions, we 
eliminate the $T_{\rm period}=2\pi/\Omega$ oscillation 
by taking the average over each period, 
as denoted by ``avg".  
The time profiles of the photo-doping rate and the 
total energy in Fig.\ref{fig:energy} are similar, which indicates that
the energy absorbed from the AC field is 
used to excite electron-hole pairs, i.e., photo-doping. 
The system relaxes to a steady state until the situation where 
pair production rate = annihilation (stimulated emission) rate
is achieved.

The doping rate depends both on the 
frequency $\Omega$ and the strength $F$ of the
electric field. When $\Omega$ is greater than the 
Mott gap $\Delta$, 
excitations occur via one-photon absorptions. 
Even when $\Omega<\Delta$, multi-photon processes 
can excite the system above the gap for sufficiently 
strong fields, 
where the process becomes resonant when $m\Omega=\Delta$ 
with $m$ being an integer \cite{OkaAokiunpublished}.
The extreme case is the DC limit $\Omega\to 0$, where
many-body Landau-Zener tunneling across the Mott gap 
induces metallization \cite{Oka2003,Oka2005a,Oka_LZreview}.


\begin{figure}[t]
\centering 
\includegraphics[width=8.5cm]{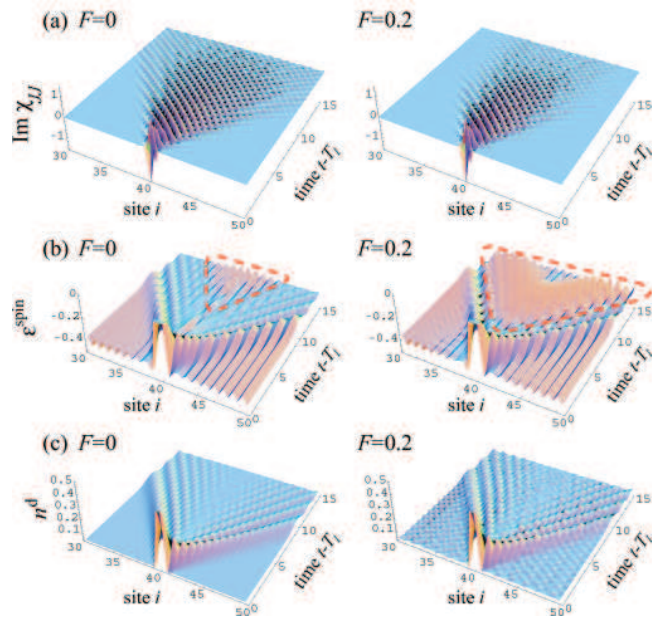}
\caption{
(Color online) 
Current-current correlation $\mbox{Im}\chi_{JJ}(t,T_1;i,j)$(a), 
the local spin energy $\ve^{\rm spin}_i(t)$(b), 
and the double occupancy $n^{\rm d}_i(t)$(c) 
for zero field $F=0$ (left panels), 
and for a finite field $F=0.2$ (right) 
with $U/t_{\rm hop}=8.0,\;\Omega/t_{\rm hop}=8.0$. 
Here the probed position $j$ in the pump-probe process is set to be 
at the center ($j=40$) of the 1d system of length 80.  
Red broken lines indicate the region for relaxation due to spin (see text).
The small ripples (c, right panel) is the $T=2\pi/\Omega$ period 
fluctuation.
}
\label{fig:correlationfunction}
\end{figure}
{\it Correlation functions ---}
To characterize the 1d many-body 
system we calculate the correlation functions 
after a steady state is attained,  
\begin{eqnarray}
\chi_{AB}(t,T_1;i,j)=\bra\Psi(t)| A_iU(t,T_1)B_j|\Psi(T_1)\ket,
\end{eqnarray}
where $A$, $B$ are operators, e.g.,
the current $J_i=-it_{\rm hop}\sum_\sigma(c_{i+1\sigma}^\dagger
c_{i\sigma}-\mbox{h.c.})$, or 
the spin $\vec{s}_i=\frac{1}{2}\sum_{\alpha\beta}
c_{i\alpha}^\dagger\vec{\sigma}_{\alpha\beta}c_{i\beta}$, and 
$T_1$ is the time around which the steady state is 
reached (and the curves in Fig.\ref{fig:energy} 
flatten; typically $T_1=50,\;100$ depending on the field strength). 
Physically, $\chi_{JJ}$ represents the probing process 
in standard pump-probe experiments, 
where a photon in the probe light generates a 
local electron-hole pair at position $j$.  
To focus on the interplay between the 
charge and spin degrees of freedom, we can compare 
in Fig. \ref{fig:correlationfunction} the current correlation function 
with the behaviors of spin and charge.  
For the spin we define the local spin energy, 
$\ve^{\rm spin}_i(t)/J_{\rm exch} \equiv
\bra\Psi^{ J}(t)|{\Vec s}_{i+1}\cdot{\Vec s}_i|\Psi^{ J}(t)\ket$, 
which measures the exchange energy 
(normalized by the exchange coupling 
$J_{\rm exch}=4t_{\rm hop}^2/U$)\cite{commentHeisenberg}.  
For the charge we examine the double occupancy defined by 
$n^{\rm d}_i(t)=\bra\Psi^{ J}(t)|n_{i\up}n_{i\dw}|\Psi^{ J}(t)\ket$. 
Here $|\Psi^{ J}(t)\ket=U(t,T_1)J_j|\Psi(T_1)\ket$ is the 
state where a perturbation ($J_j$) is added at site $j$ on the 
steady, nonequilibrium state.  
The behavior of the three quantities shows that the temporal evolution 
after the probe-excitation on $j$ at $t=T_1$ propagates in two processes. 
The first is {\it diffusion of the doublon-hole pair}, 
which is followed by the {\it relaxation} process.  
The relaxation is seen as a decay of the 
current correlation, which is seen to be accompanied by 
a disturbance in the spin structure 
(as marked with a red broken line in the figure).  
This indicates that the spin and charge degrees of 
freedom become coupled more strongly, where spins act as a kind of energy 
reservoir for charges.  
Spin-charge coupling already exists 
in equilibrium for higher-energy states,\cite{KusakabeAoki91} 
which is natural since charge excitations act as boundary conditions to spins 
with spin vanishing at doubly occupied or empty sites. 
The decay of the current correlation, already present for zero AC electric field ($F=0$), becomes faster in finite fields, which implies that the 
spin-charge coupling becomes stronger, due to higher-energy states become 
involved.  
However, in the field range studied here, the coupling is not strong 
enough to destroy the spin-charge separation picture \cite{Takahashi08},
i.e.,
the spin and charge degrees of freedoms still have 
independent dispersions, as discussed below.

\begin{figure}[t]
\centering 
\includegraphics[width=7.2cm]{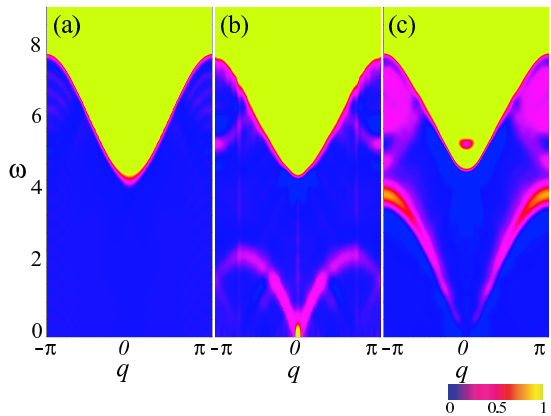}
\caption{
(Color online) 
The optical spectrum, $\mbox{Im}\chi_{JJ}(q,\omega)$, 
with $U/t_{\rm hop}=8.0$ for 
(a) the half-filled case with no AC fields, 
(b) the half-filled with a finite AC field $F/t_{\rm hop}=0.1$, 
and (c) the electron-doped with no AC fields ($n=1.025,\;F=0$).
}
\label{fig:spec}
\end{figure}
{\it Collective excitations in nonequilibrium ---} 
So what is the nature of the photo-induced carriers?  
We can obtain the excitation spectrum as 
the Fourier transform of the correlation functions,
\begin{eqnarray}
&&\chi_{AB}(q,\omega)=\\
&&\int_{T_1}^{T_1+T_2} dt\sum_j e^{i\omega (t-T_1)-iq(j-j_{c})}
\mbox{Im}
\chi_{AB}(t,T_1;j,j_c),\nonumber
\end{eqnarray}
where $T_2$ is set to be prior to the time at which the wavefront reaches 
the sample boundary.    
We call the quantity, following the equilibrium case, 
the optical spectrum for $\chi_{JJ}$, and 
the spin spectrum for $\chi_{ss}$.

A striking feature in the result in Fig. \ref{fig:spec}, 
a first key finding here, is that, 
while we have an optical gap (= Mott gap) in the zero AC field 
(Fig. \ref{fig:spec}(a)), a set of new states with a 
zero-gap, linear dispersion,
$\omega_q\sim v_{\rm PTL}|q|,$ 
emerges in the gap in finite AC fields 
(Fig. \ref{fig:spec}(b),(c)).  
We call the metallic state a 
{\it photo-induced Tomonaga-Luttinger-like liquid} 
in the sense that its charge velocity $v_{\rm PTL}$ 
is renormalized by the electron interaction.  
Its value is in fact similar to the 
equilibrium counterpart as can be seen by comparing the slope 
in nonequilibrium with that for the equilibrium, doped system 
(Fig. \ref{fig:spec}(c)).

\begin{figure}[th]
\centering 
\includegraphics[width=8.5cm]{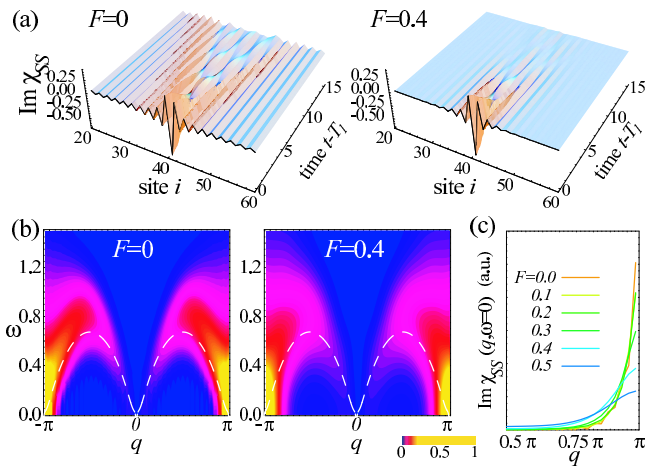}
\caption{
(Color online) 
(a) Spin correlation $\mbox{Im}\chi_{SS}
(t,T_1;i,j) $
for zero field (left) and for a finite
field $F=0.4$ (right) with 
$U/t_{\rm hop}=8.0,\;\Omega/t_{\rm hop}=8.0$.
(b) Color-coded spin spectrum $\mbox{Im}\chi_{ss}(q,\omega)$
in the half-filled Hubbard model with $U/t_{\rm hop}=8.0$ 
for zero AC field(left), or for a finite AC field $F/t_{\rm hop}=0.4$(right). 
The white dashed lines are the des Cloiseaux-Pearson mode.
(c) The AF peak $\mbox{Im}\chi_{ss}(q,\omega=0)$ vs $q$ are also displayed for 
various field strengths.
}
\label{fig:spin}
\end{figure}

If we turn to the spin spectrum in Fig. \ref{fig:spin}, 
we can clearly see the 
des Cloizeaux-Pearson mode 
in equilibrium (precisely speaking, this was obtained by the Bethe ansatz method\cite{Woynarovich1983}), 
which survives in finite fields 
. 
We note that the charge and spin velocities defined and numerically 
obtained in nonequilibrium here are different. So in this particular sense 
we have a spin-charge separation.
As the field becomes stronger, the 
spectrum becomes blurred.   
Specifically, the antiferromagnetic fluctuation represented
by the peak value $\chi_{ss}(q\sim\pi,\omega\sim 0)$,
becomes smaller, which should be because the induced carriers 
act to melt the magnetic order.

{\it An effective model ---}
Due to its many-body nature, it is difficult to obtain 
a mathematically rigorous analysis of the photo-induced 
metallic state in Mott insulators.  
However, we can carve out some of its properties
in terms of an effective model, 
which was initiated by Luther, Emery, and by Giamarchi
to study linear-responses 
in the presence of a charge gap \cite{LutherEmergy74,Emery90,Giamarchi1991,Mori1996}. 
In this approach we start from the 1+1 dimensional 
massive Thirring model. 
The charge degrees of freedom is represented by
two spinless fermions $\Psi=(\psi_1(x), \psi_2(x))$, 
where $\psi_1, \psi_2$ represent left mover and right mover, respectively, 
and the Hamiltonian reads
\begin{eqnarray}
H_{\rho}^{\rm MT}(t)=v_c\int dx
\left[\Psi^\dagger(x)(-i\nabla_x(t)\sigma_3+\frac{\Delta}{2}\sigma_1)\Psi(x)\right]
+H_{{\it I}},
\label{MT}
\end{eqnarray}
where $\sigma_i$ is Pauli matrices, $v_c$ the renormalized charge 
velocity, and $\Delta$ the Umklapp-scattering coupling constant, 
an ascendant of the original Mott gap at half filling. 
This model is dual to the sine-Gordon model, where the 
size of the interaction term 
$H_{{\it I}}=g\int dx[(\Psi^\dagger\Psi)^2-(\Psi^\dagger
\tau_1\Psi)^2$ 
translates to the sine-Gordon coupling through a 
duality relation\cite{Coleman:1976uz}.  
However, in the following we make a 
further simplification, namely, we 
neglect the self-interaction $H_I$. 
This by no means implies a neglect of the original 
electron-electron interaction, but amounts to 
neglect self-energy corrections to quasi-particle 
life time, etc.  We employ this approximation to 
focus on the effect of the charge gap, which primarily 
appears from the first term in eqn.(\ref{MT}).  

The AC electric field is taken as the coupling, 
$\nabla_x(t)=\pa_x+iA_1(t)$, in the above, with
$A_1(t)=(F/\Omega)\sin\Omega t$.  
We then obtain the nonlinear, nonequilibrium evolution of the 
state in this model.  After a Fourier transform, 
the equation of motion becomes 
$i\frac{d}{dt}|\Psi_k(t)\ket
=\left[v_c(k+(F/\Omega)\sin\Omega t)
\sigma_3+\frac{\Delta}{2}\sigma_1)\right]|\Psi_k(t)\ket$. 
Here we adopt the Floquet method for treating 
systems in AC fields (see e.g., \cite{HanggiREVIEW1998}), 
i.e., we seek a solution of the form 
$|u_\alpha(k;t)\ket=
\sum_me^{-i\ve_\alpha(k) t-im\Omega t}|u^m_\alpha(k)\ket$,
where $m$ is the Floquet index, $\alpha$ 
labels the eigenstate, and $\ve_\alpha$ is Floquet's quasi-energy.
From the equation of motion, 
the Floquet modes satisfy a set of linear relations 
$\sum_{n}(H_{\rm eff})^{mn}|u^{n}_\alpha(k)\ket=
\ve_\alpha(k)|u^{m}_\alpha(k)\ket$,
where the effective Floquet Hamiltonian has matrix elements 
$
(H_{\rm eff})^{mm}=v_ck\sigma_3+\frac{\Delta}{2}\sigma_1-m\Omega\sigma_0,\;
(H_{\rm eff})^{m,m\pm 1}=\mp \frac{i}{2}v_c\frac{F}{\Omega}\sigma_3$.
We consider a sudden switch on of the AC field, 
i.e., the time evolution starts from 
the groundstate $|\psi_0(k)\ket$ of 
$v_ck\sigma_3+\frac{\Delta}{2}\sigma_1$. 
The solution is then $
|\psi_0(k;t)\ket=\sum_\alpha\phi_\alpha(k)|u_\alpha(k;t)\ket$
with $\phi_\alpha(k)=\bra u^{0}_\alpha(k)|\psi_0(k)\ket$.
\begin{figure}[t]
\centering 
\includegraphics[width=8.cm]{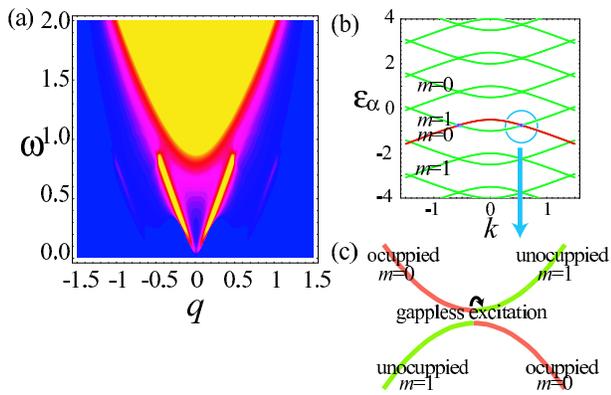}
\caption{
(Color online) 
(a) The optical spectrum $\mbox{Im}\chi^{\rm Dirac}(q,\omega)$ of the effective Dirac model in a finite AC field 
with  $\Omega/\Delta=1.5,\;F=0.1$.  
(b) The Floquet quasi-energy $\ve_\alpha(q)$, 
where the red lines represent 
occupied states, green lines unoccupied states.
(c) A schematic level repulsion.
}
\label{fig:qed}
\end{figure}
The current correlation function in this model is \cite{Oka2008gr}
\begin{eqnarray}
&&\mbox{Im}\chi^{\rm Dirac}(q,\omega)=2 e^2\int\frac{dk}{2\pi}\nonumber\\
&&\times \sum_{\alpha\beta m}(|\phi_\beta(k)|^2-
|\phi_\alpha(k+q)|^2)
\delta(\ve_\beta(k)-\ve_\alpha(k+q)+\omega)\nonumber\\
&&
\times
\mbox{Tr}\left[|u_\alpha^0 (k+q)\ket\bra u_\alpha^m(k+q)|\tau_3
|u_\beta^m(k)\ket\bra u_\beta^0(k)|\tau_3\right].
\label{eq:ImChiDirac}
\end{eqnarray}
The optical spectrum, Fig.\ref{fig:qed}(a), shows that a metallic, 
linear-dispersion mode emerges in the gap in the presence of an AC field.  
The result does resemble the present td-DMRG result (Fig.\ref{fig:spec}). 
The spectrum, despite being nonequilibrium, is also similar to 
the massive Thirring model result for the doped 1d Mott insulator.\cite{Mori1996} 
The origin of the gapless excitation in nonequilibrium 
can be traced back, in the present massive Thirring + Floquet analysis, 
to the quasi-energy level scheme in Fig. \ref{fig:qed}(b). 
In the absence of AC fields, the Dirac model has 
two branches with the lower band completely filled. 
As we turn on the AC field, we have a series of 
replicas (i.e., Floquet modes equally spaced by $\Omega$) 
for each of the electron and hole branches. 
Physically, these modes correspond to 
$m$-photon absorbed states. As in standard quantum mechanics, 
level repulsion takes place when two 
modes cross with each other with nonzero matrix element between them.  
In our model, the most important feature is the 
level repulsion between an occupied Floquet mode 
(red in Fig.\ref{fig:qed}(b)) and an unoccupied one (green). 
As shown in the blowup (Fig.\ref{fig:qed}(c)), 
gapless excitations then emerge across the 
occupied and unoccupied states, which should contribute to the 
optical spectrum when we evaluate eqn.(\ref{eq:ImChiDirac}).  
Specifically, the linear dispersion of the photo-induced state is $\omega_q\sim v_c|q|$, where the renormalized velocity $v_c$, a parameter independent of the spin velocity $v_s\propto t_{\rm hop}^2/U$, is conceived to have a value that depends on the detail of the irradiation.


Another interesting point 
is when the photon energy $\Omega$ is below the gap $\Delta$. 
In this situation, multi-photon processes with $m\geq 2$ are necessary 
to excite the system. We can indeed show that, 
when the condition $\Omega\sim \Delta/m$ is 
met, the excitation becomes resonant and carriers are
injected efficiently, which will be described elsewhere.

In conclusion, we have shown that 
a Mott insulator irradiated by 
strong AC electric fields has a collective mode 
which is reminiscent of the 
Tomonaga-Luttinger liquid in equilibrium. 
While spins and charges are coupled in the relaxation 
process, the collective modes have different 
spin and charge velocities. 
Emergence of the linear collective mode is also 
supported by an effective Dirac model.  
We have greatly benefited from discussions with Naoto Tsuji
on the Floquet method. TO acknowledges
Keiji Saito and Keiichiro Nasu for valuable discussions.  
This work has been supported in part by a Grant-in-Aid for Scientific Research 
on a Priority Area ``Anomalous quantum materials" 
from the Japanese Ministry of Education.


\end{document}